\documentclass{article}
\usepackage{spconf,amsmath,graphicx,amssymb}
\usepackage{algorithm}
\usepackage{algpseudocode}
\usepackage{color}
\usepackage{cite}

\title{PRIVACY-COST TRADE-OFF IN A SMART METER SYSTEM WITH A RENEWABLE ENERGY SOURCE AND A RECHARGEABLE BATTERY}
\name{ Ecenaz Erdemir, Pier Luigi Dragotti and Deniz G{\"u}nd{\"u}z \thanks{This work has been funded by the European Research Council (ERC) through project BEACON (No. 725731), and the UK EPSRC through project COPES (EP/N021738/1.}}
\address{Department of Electrical and Electronic Engineering, Imperial College London, UK}
\begin{document}
%
\maketitle
\begin{abstract}
\label{sec:abstract}

\vspace{-0.2cm}
We study the privacy-cost trade-off in a smart meter (SM) system with a renewable energy source (RES) and a finite-capacity rechargeable battery (RB). Privacy is measured by the mutual information rate between the energy demand and the energy received from the grid, where the latter also determines the cost, and hence, reported by the SM to the utility provider (UP). We consider a renewable energy generation process that fully charges the RB at random time instants, and its realization is assumed to be known also by the UP. We reformulate the problem as a Markov decision process (MDP), and solve it by dynamic programming (DP) to design battery charging and discharging policies that minimize a linear combination of the privacy leakage and energy cost. We also propose a lower bound and two alternative low-complexity energy management policies, one of which is shown numerically to perform close to the MDP solution.
\end{abstract}
\vspace{-0.5cm}
\section{Introduction}

\vspace{-0.3cm}
\label{sec:intro}
Smart meters (SMs) are essential components of smart grids: they collect real-time consumption data of a household, and report it to the utility provider (UP). SM measurements can be used for time-of-use pricing, trading user-generated energy, and mitigating load variations \cite{DGGiulioVPoor}. However, SM readings can also reveal details about consumer's private activities, which they may not want to share with the UP. Various techniques have been proposed in the literature to enable SM privacy \cite{Encrypt,Noise,SMviaTrapdoor,UfukTopcuRB,wLoss,Khisti,ParvTO,GiulioFB,GiulioIB,Gomez:TIFS:13,TanGunduzPoor,EMU_ctrl}, which can be categorized as those based on SM data manipulation, and those based on demand shaping. While the former focuses on modifying SM measurements \cite{Encrypt,Noise}, the latter directly manipulates energy consumption exploiting physical resources, such as a rechargeable battery (RB) \cite{SMviaTrapdoor,UfukTopcuRB,wLoss,Khisti,ParvTO} or a renewable energy source (RES) \cite{GiulioFB,GiulioIB,Gomez:TIFS:13,TanGunduzPoor,EMU_ctrl}. Manipulating the SM readings reduces the relevance of the reported vaues for grid management and load prediction, limiting the benefits of SMs. Moreover, the grid operator can place sensors outside a household and obtain the real consumption data, since they own and control the infrastructure. On the other hand, demand shaping tackles these issues by manipulating the real consumption. In demand shaping, instantaneous demand of a user can be supplied partially by the power grid, as the rest can be provided by the RB or RES. This effectively filters the energy consumption time series, reducing the information leakage to the UP.

In \cite{Khisti}, information theoretic privacy with a RB is formulated as a Markov decision process (MDP). Markovian energy demand is considered, and the minimum leakage is obtained numerically through dynamic programming (DP), while a single-letter expression is obtained for an independent and identically distributed (i.i.d) demand. This approach is extended to the scenario with a RES in \cite{GiulioFB}. In \cite{ParvTO}, privacy-cost trade-off is examined with a RB. Due to Markovian demand and price processes, the problem is formulated as a partially observable MDP with belief-dependent rewards ($\rho$-POMDP), and solved by DP for infinite-horizon.

We consider the privacy-cost trade-off of a SM system with both a RB and a RES \cite{ParvTO, Tan:TIFS:17}. We measure privacy as the mutual information rate between the user demand and the energy received from the grid. We define the cost as the average amount of energy received from the grid, and study the trade-off between the cost and privacy by setting their weighted sum as the objective function. We formulate the problem as a MDP with a continuous belief state, and solve it numerically by DP by quantizing the belief state. Our contribution with respect to \cite{ParvTO} is the inclusion of a RES into the system, which provides additional privacy.
While \cite{GiulioFB} also considers RB and RES jointly, here we study the privacy-cost trade-off, and present numerical solutions focusing on a particular renewable energy arrival process that fully recharges the RB at random time instances. We also provide two low-complexity policies and a lower bound, which exploit the special structure of the renewable energy process. We show numerically that the policy targeting a fixed recharge period performs very close to the infinite-horizon MDP solution, providing a low-complexity alternative for practical systems.

\vspace{-0.6cm}
\section{System Model}

\begin{figure}[tp]
\centering
\includegraphics[scale=0.4]{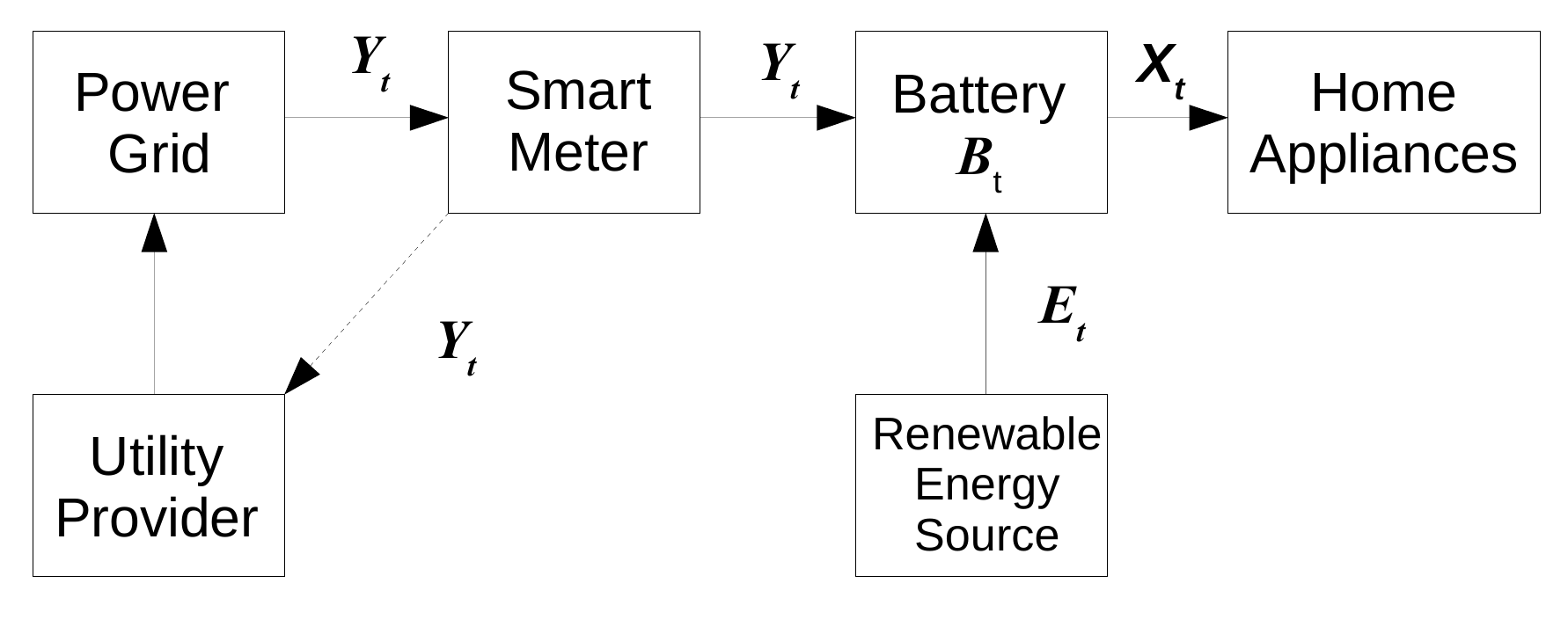}
\vspace{-0.3cm}
\caption{Illustration of the SM system model.} 
\label{System_Model1}
\end{figure}

\vspace{-0.4cm}
\label{sec:SystemModel}
 We consider a discrete time system model, illustrated in Fig.~\ref{System_Model1}, in which the energy demand of the user and energy requested from the grid at time slot $t$ are denoted by $X_t \in \mathcal{X}$ and $Y_t \in \mathcal{Y}$, respectively, where $(|\mathcal{X}|,|\mathcal{Y}| < \infty)$. The RB state of charge at the beginning of time slot $t$ is denoted by $B_t \in \mathcal{B}$:=$\{0, \dots, B_{max}\}$. The RB charging and discharging process is assumed ideal without any losses (see \cite{wLoss} for a model with energy losses). $E_t \in \mathcal{E}$:=$\{0,B_{max}\}$ units of energy is generated by the RES at the beginning of each time slot $t$; that is, when the renewable energy arrives, it completely recharges the RB, and it can be used by the appliances only through the RB. The $E_t$ process is assumed to be independent of $X_t$, and known by the UP.
We assume that $X_t$ and $E_t$ are i.i.d. with distributions $P_X$ and $P_E$, respectively.

The appliances' energy demand is always satisfied; that is, $E_t+B_t+Y_t \geq X_t$, $\forall t$. 
In addition, intentional energy waste to provide privacy, or selling energy to the grid are not allowed. Therefore, the battery state of charge is updated as

\vspace{-0.7cm}
\begin{align}\label{eqn:BatteryUpdate}
B_{t+1}=\text{min}(E_t+B_t,B_{max})+Y_t-X_t, \hspace{0.5cm} \forall t,
\end{align}
where $Y_t$ is chosen such that $B_{t+1} \leq B_{max}$. 

The amount of energy requested from the grid is determined by a randomized battery charging policy $\boldsymbol{q}=\{q_t\}^{\infty}_{t=1}$, where $q_t$ is a conditional probability distribution $q_t(Y_t| X^t,  B^t,E^t, Y^{t-1})$, which randomly decides on the amount of energy received from the grid at time $t$, given the histories of demand $X^t$:=$\{X_1, \dots, X_t\}$, battery charge $B^t$, energy generation $E^{t}$, and grid energy $Y^{t-1}$. 
Our goal is to find an energy management policy, $\{q^*_t\}^{\infty}_{t=1}$, which provides the best trade-off between the privacy and cost. 

\vspace{-0.5cm}
\subsection{Privacy Measure}

\vspace{-0.2cm}
\label{ssec:PrivacyCostMeasure}
We measure the privacy of a policy $\boldsymbol{q}=\{q_t\}^{T}_{t=1}$ over $T$  time slots by the \textit{information leakage rate}, $L_{\boldsymbol{q}}(T)$, defined as the average mutual information between the demand side load $X^T$ and initial RB charge $B_1$, and the SM readings $Y^T$:
\vspace{-0.3cm}
\begin{align}
\label{eqn:PrivacyMeasure}
L_{\boldsymbol{q}}(T) := \frac{1}{T}I(X^T,B_1;Y^T|E^T),
\end{align}
where $E^T$ is known by the UP. It can be shown, similarly to \cite{Khisti}, that there is no loss of optimality in considering policies of the form $q_t(Y_t| X_t, B_t,E^t, Y^{t-1})$; that is, it is sufficient to consider only the current demand and battery state. Hence, (\ref{eqn:PrivacyMeasure}) can be rewritten in an additive form 

\vspace{-0.7cm}
\begin{align}\label{eqn:AdditiveLeakage}
L_{\boldsymbol{q}}(T) = \frac{1}{T} \sum_{t=1}^{T} I(X_t,B_t;E_t,Y_t|Y^{t-1},E^{t-1}).
\end{align}

\vspace{-0.3cm}
The Markovity of optimal actions and the additive objective function in (\ref{eqn:AdditiveLeakage}) allow us to represent the privacy component of our problem as a MDP with state $S_t$=$\{B_t, X_t\} \in \mathcal{S}$. However, the leakage at time $t$ depends on $Y^{t-1}$ and $E^{t-1}$, resulting in a growing state space in time. Therefore, a belief state $\beta_t(s_t)$ is defined as the causal posterior probability distribution over the state space given $Y^{t-1}$ and $E^{t-1}$:

\vspace{-0.7cm}
\begin{align}\label{eqn:BeliefState}
\beta_t(s_t)=P^{\boldsymbol{q}}(S_t=s_t|Y^{t-1}=y^{t-1},E^{t-1}=e^{t-1}).
\end{align}

\vspace{-0.3cm}
The control actions chosen by randomized policies are the conditional probabilities of energy received from the grid given the state and belief, and denoted by $a_t(y_t|s_t,e_t)=P^{\boldsymbol{q}}(Y_t=y_t|S_t$=$s_t,E_t$=$e_t,\beta_t)$ \cite{GiulioFB}.

We follow the approach in \cite{GiulioFB} for updating the belief state, and define the per-step leakage of taking action $a_t(y_t|s_t,e_t)$ which is incurred by the policy $\boldsymbol{q}$ at each step as,

\vspace{-0.55cm}
\begin{align} 
\label{eqn:PerStepObjective}
 l_t(s_t,e^t,a_t,y^t;\boldsymbol{q}) := \log \frac{a_t(y_t|s_t,e_t)P_E}{P^{\boldsymbol{q}}(y_t,e_t|y^{t-1},e^{t-1})}.
\end{align}

\vspace{-0.25cm}
\sloppy
The average leakage rate over a finite-horizon $T$, $\frac{1}{T}\mathbb{E}_{\boldsymbol{q}}[\sum_{t=1}^{T}l_t(s_t,e^t,a_t,y^t) ]$, is equal to the original formulation in (\ref{eqn:AdditiveLeakage}). Given belief and action probabilities, average information leakage rate at time $t$ is formulated as,

\vspace{-0.7cm}
\begin{align}
\mathbb{E}_{\boldsymbol{q}}[l_t(s_t,e^t,a_t,y^t)&]=\sum_{ \substack{s_t \in \boldsymbol{S},e_t \in \boldsymbol{E}\\  y_t \in \boldsymbol{Y}}} \beta_t(s_t)a_t(y_t|s_t,e_t)P_E \nonumber \\  & \times \log \frac{a_t(y_t|s_t,e_t)P_E}{\sum\limits_{\substack{\hat{s}_t \in \boldsymbol{S}}} \beta_t(\hat{s}_t)a_t(y_t|\hat{s}_t,\hat{e}_t)P_E}\nonumber \\
&:=\mathcal{L}(\beta_t,a_t).
\label{eqn:AvgLeakageRate}
\end{align}

\vspace{-0.9cm}
\subsection{Energy Cost}

\label{ssec:CostMes}
\vspace{-0.3cm}
Energy cost is defined as the average amount of energy received from the grid over $T$ time slots,

\vspace{-0.7cm}
\begin{align}
\label{eqn:GridPurchaseCost}
C_{\boldsymbol{q}}(T) := \frac{1}{T} \sum \limits^{T}_{t=1}Y_t.
\end{align}

\vspace{-0.3cm}
We remark that, differently from \cite{ParvTO}, we do not consider time-varying energy unit cost, although our model can easily be extended in this direction. In the context of \cite{ParvTO}, i.e., in the absence of a RES, our cost model would result in a deterministic energy cost, independent of policy $q_t$. However, in the presence of a RES, our cost model follows \cite{TanGunduzPoor}, and incentivizes the maximum exploitation of locally generated renewable energy. For example, when privacy is not a concern, cost minimizing policy would use battery energy first, to be able to store the arriving renewable energy as much as possible. 
The energy cost averaged over a finite-horizon $T$ is simply $\frac{1}{T} \mathbb{E}_{\boldsymbol{q}}[\sum_{t=1}^{T} y_t]$.
The average per-step cost can be represented in terms of belief and action probabilities as follows:

\vspace{-0.6cm}
\begin{align}
\mathcal{C}(\beta_t,a_t) := \mathbb{E}_{\boldsymbol{q}}[y_t] = \sum_{\substack{s_t \in \boldsymbol{S}, e_t \in \boldsymbol{E},\\ y_t \in \boldsymbol{Y}}} \beta_t(s_t)P_Ea_t(y_t|s_t,e_t) y_t \nonumber
\end{align}

\vspace{-0.75cm}
\subsection{Weighted Total Privacy Leakage and Energy Cost}

\vspace{-0.2cm}
We have two distinct performance measures, which are not necessarily aligned. Therefore, we define the objective function as the weighted sum of the information leakage rate and the average cost over all the feasible policies, as $T \rightarrow \infty$,

\vspace{-0.65cm}
\begin{align}\label{eqn:WeightedObjective}
 U^*(\gamma) = \lim_{T \rightarrow \infty}\inf_{\boldsymbol{q}} [ \gamma L_{\boldsymbol{q}}(T) + (1-\gamma) C_{\boldsymbol{q}}(T) ],
\end{align}
where $ \gamma \in [0,1]$ is determined by the user according to her preference regarding privacy and cost. Our goal is to design a policy $\boldsymbol{q}^*(\gamma)$, which is the $\mathrm{argmin}$ of the right hand side of (\ref{eqn:WeightedObjective}), satisfying the energy management rules. This problem can be modeled as an MDP with state $\beta_t(s_t)$ and action $a_t(y_t|s_t,e_t)$. The corresponding Bellman equations can be written similarly to \cite{Khisti} and \cite{GiulioFB}. We include the instantaneous weighted objective function, $\mathcal{U}_{\gamma}(\beta_t,a_t) = [\gamma \mathcal{L}(\beta_t,a_t) + (1-\gamma) \mathcal{C}(\beta_t,a_t)]$, into the Bellman operator,

\vspace{-0.6cm}
\begin{align}
\hspace{-0.3cm}
[T_av](\beta)\text{ = }\mathcal{U}_{\gamma}(\beta,a)\text{ +} \hspace{-0.4cm}
\sum_{\substack{s \in S, e \in E \\ y \in Y}} \hspace{-0.3cm} \beta(s)a(y|s,e)P_Ev(\phi (\beta,y,a)), \label{eqn:Bellman}
\end{align}
where $v$ is the value function and the updated belief state is represented by $\beta_{t+1}=\phi(\beta_t,y_t,a_t)$. The mplementation of DP for infinite-horizon is as follows:

\vspace{-0.4cm}
\begin{itemize}
    \item For $\lambda$ constant \cite{Puterman}, the value function $v$ is time-homogeneous and defined iteratively:
 \vspace{-0.3cm}   
\begin{align}
\lambda +v(\beta)= \min_{a}[T_av](\beta).
\end{align}

\vspace{-0.6cm}
    \item Time-homogeneous optimal policy, $\textbf{q}^*=(q^*, q^*, \dots)$,
    
 \vspace{-0.8cm}   
\begin{align}
q^*(y_t|s_t,e_t,\beta_t)=a(y_t|s_t,e_t).
\end{align}
\end{itemize}

\vspace{-0.4cm}
While an exact DP solution cannot be achieved due to the continuous belief, we provide an approximate numerical solution. To be able to solve the problem numerically by DP, we discretize the belief $\beta_t(s_t)$. At each value iteration, we quantize the updated belief, $\beta_{t+1}(s_{t+1})$, by rounding it to the closest discrete belief value.

\vspace{-0.5cm}
\section{Low-Complexity Policies}

\vspace{-0.35cm}
Due to the special renewable energy generation process we consider here, the problem is an episodic MDP, which resets to an initial state of full RB at every renewable energy instant. Between two consecutive energy arrivals, energy transitions occur only between the grid, RB and home appliances. Hence, for each time period between two charging instances, the system can be modeled as a SM with only a RB and no RES. Accordingly, we formulate a finite-horizon privacy-cost trade-off problem for a SM system with an initially full RB, which will be used to propose a low-complexity policy as well as a lower bound  for the original problem. 

In the finite-horizon problem with no RES, as before, the user demand is always satisfied by imposing $Y_t+B_t \geq X_t$, $\forall t$, and RB charge is updated by $B_{t+1}$=$B_t+Y_t-X_t,  \forall t$. Randomized battery charging policies, $q_t$, are of the form $q_t(Y_t| X_t, B_t, Y^{t-1})$. The information leakage rate induced by the policy $\boldsymbol{q}$ over a finite-horizon between two consecutive energy arrivals is given by,

\vspace{-0.7cm}
\begin{align}\label{eqn:PrivacyMeasureSub}
\Bar{L}_{\boldsymbol{q}}(T) := \frac{1}{T}I(X^T,B_1;Y^T).
\end{align}

\vspace{-0.3cm}
As in Section \ref{sec:SystemModel}, (\ref{eqn:PrivacyMeasureSub}) can be written in an additive form:

\vspace{-0.7cm}
\begin{align}\label{eqn:AdditiveLeakageSub}
\Bar{L}_{\boldsymbol{q}}(T) = \frac{1}{T} \sum_{t=1}^{T} I(X_t,B_t;Y_t|Y^{t-1}).
\end{align}

\vspace{-0.3cm}
Similarly to the original problem, the finite-horizon problem with no RES can also be formulated as a MDP with belief $\beta_t(b_t,x_t)$. Control actions used to determine energy received from the grid are defined as $a_t(y_t|b_t,x_t)=P^{\boldsymbol{q}}(Y_t$=$y_t|B_t$=$b_t,X_t$=$x_t,\beta_t)$.  To solve the finite-horizon problem by DP, we use the method in \cite{Khisti} for belief updates, and express the average information leakage rate at time $t$ in terms of belief and action probabilities as follows:

 \vspace{-0.85cm}
\begin{align}
\mathbb{E}_{\boldsymbol{q}}[\Bar{l}_t(b_t,x_t,a_t,y^t;\boldsymbol{q})&]=\hspace{-0.2cm} \sum_{ \substack{b_t \in \boldsymbol{B},x_t \in \boldsymbol{X}\\  y_t \in \boldsymbol{Y}}} \hspace{-0.2cm} \beta_t(b_t,x_t)a_t(y_t|b_t,x_t) \nonumber \\  & \times \log \frac{a_t(y_t|b_t,x_t)}{\hspace{-0.2cm} \sum\limits_{\substack{\hat{b}_t \in \boldsymbol{B}, \hat{x}_t \in \boldsymbol{X}}} \hspace{-0.2cm} \beta_t(\hat{b}_t,\hat{x}_t)a_t(y_t|\hat{b}_t,\hat{x}_t)}\nonumber \\
&:=\mathcal{\Bar{L}}(\beta_t,a_t).
\label{eqn:AvgLeakageRateSub}
\end{align}

\vspace{-0.4cm}
The average per-step energy cost for the finite-horizon problem is determined similarly to Section \ref{ssec:CostMes}, and represented in terms of the belief and action probabilities as:

\vspace{-0.8cm}
\begin{align}
\label{eqn:PerStepCostSub}
 \mathcal{\Bar{C}}(\beta_t,a_t) := \mathbb{E}_{\boldsymbol{q}}[y_t] = \hspace{-0.4cm} \sum_{\substack{b_t \in \boldsymbol{B}, x_t \in \boldsymbol{X} \\ y_t \in \boldsymbol{Y}}} \hspace{-0.4cm} \beta_t(b_t,x_t)a_t(y_t|b_t,x_t) y_t.
\end{align}

\vspace{-0.4cm}
The weighted objective function to be minimized for the finite-horizon privacy-cost trade-off is denoted by 
$\mathcal{\Bar{U}}_{\gamma}(\beta_t,a_t)$=$[\gamma \mathcal{\Bar{L}}(\beta_t,a_t)$+$(1-\gamma) \mathcal{\Bar{C}}(\beta_t,a_t)]$. We can first quantize the belief state, and solve the resulting MDP with a finite state space by DP recursively using the Bellman operator in (\ref{eqn:Bellman}) with the corresponding changes for finite-horizon.

In the next subsection, using the solution of the finite-horizon problem above, we will propose a low-complexity solution for the original infinite-horizon problem with an RB and RES, and a resetting energy generation process of the form $E_t=\{0,B_{max}\}$.

\vspace{-0.7cm}
\subsection{Threshold Policy (TP)}

\vspace{-0.3cm}
\label{ssec:ThresholdPolicy}
In TP, we fix a target horizon $n$, and after each RB recharge instance, start employing the optimal energy management policy for this finite-horizon, derived in the previous section. We follow the optimal policy for horizon $n$ until either the battery is recharged again, in which case we restart with the same policy, or we reach the time horizon $n$. If the RB is not recharged at time $(n+1)$, we assume that we simply provide all the energy demand directly from the grid, resulting in full information leakage. The intuition behind this scheme follows from the law of large numbers, which suggests that the RB will be charged after $n=\frac{1}{P_E}$ time slots with high probability. We will consider policies with a fixed time horizon of $n=\frac{1}{P_E}$, as well as those with an optimized time horizon. Our numerical results in Section 5 show that the performance with optimized but fixed time horizon closely follows that of the infinite-horizon solution.

\vspace{-0.7cm}
\subsection{Battery Conditioned Policy (BCP)}

\vspace{-0.3cm}
We propose another low-complexity policy, which depends only on the current input load. In BCP, when there is no demand, we allow the RB to be recharged by the grid with a probability $P_{C_i}$ for each battery state $B_t$=$i$, for $i$=$\{0,\dots,B_{max}\}$. On the other hand, when there is energy demand, the RB is discharged with a probability $P_{D_i}$ for each battery state. As before, intentional energy waste is not allowed. When there is demand in the case of an empty RB, it is entirely supplied from the grid. We choose $(P_{C_i},P_{D_i})$ values that minimize (\ref{eqn:AdditiveLeakage}) by an exhaustive grid search on $[0,1]^2$.

\vspace{-0.6cm}
\section{LOWER BOUND}

\begin{figure}[h!]
\centering
\includegraphics[width=8.5cm,height=2.3cm]{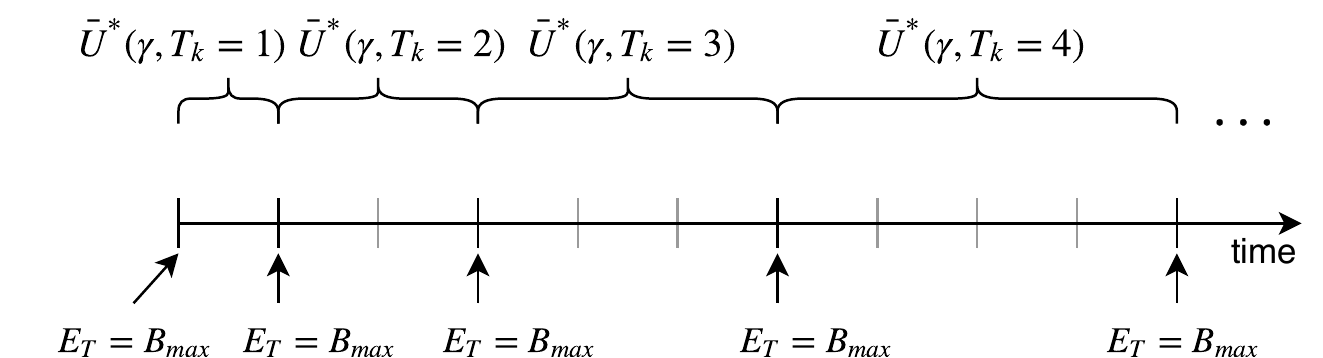}
\vspace{-0.35cm}
\caption{Renewable energy generation instances and privacy-cost rate for the corresponding intervals.} 
\label{fig:LB}
\end{figure}

\vspace{-0.4cm}
Next, we provide a lower bound on the privacy-cost trade-off by assuming that the user non-causally knows the times at which the RES recharges the RB. In Fig. \ref{fig:LB}, these time instances are represented by consecutive arrows. The weighted sum of finite-horizon leakage rate and average energy cost, minimized over policy $\boldsymbol{q}$, is denoted by $\Bar{U}^*(\gamma,T_k)$ in Fig. \ref{fig:LB}. Given i.i.d. $P_E$, the probability that the RB is recharged after $T_k$ time slots is given by
\vspace{-0.25cm}
\begin{align}
\label{eqn:fTPe}
f(T_k;P_E) = P_E(1-P_E)^{T_k}.
\end{align}

\vspace{-0.3cm}
If the RB recharge instances are known in advance, the problem reduces to the finite-horizon MDP for each inter-arrival period, and can be solved as outlined in Section 3.

\vspace{-0.4cm}

\vspace{-0.4cm}
Once the optimal performance is evaluated for all $T_k>0$, the lower bound can be derived by taking their average using the probability mass function in (\ref{eqn:fTPe}):

\vspace{-0.7cm}
\begin{align}
\label{eqn:LB}
{F}_{\gamma}(P_E)=\sum_{k=1}^{\infty}f(T_k;P_E)\Bar{U}^*(\gamma,T_k),
\end{align}
where the coefficient $f(T_k;P_E)$ approaches zero as $T$ $\rightarrow$ $\infty$, while  $\Bar{U}^*(\gamma,T_k)$ approaches the infinite-horizon privacy-cost trade-off. For the numerical solution of the infinite-sum indicated in (\ref{eqn:LB}), we perform the summation for finite $k$=$\{1,\dots,K\}$ such that $\sum^{\infty}_{k=K+1}\{f(T_{k};P_E)\Bar{U}^*(\gamma,T_{k})\} <\epsilon$. To obtain the minimum $K$ satisfying this inequality, we first consider the worst case information leakage rate and average energy cost, where all the demand is supplied by the grid, $Y_t=X_t$, and denote the lower bound by

\vspace{-0.6cm}
\begin{align*}
\label{eqn:LBconvergence}
 {F}_{\gamma}(P_E) \hspace{-0.1cm} \leq \hspace{-0.1cm} \sum_{k=1}^K \hspace{-0.05cm} f(T_k;P_E)\Bar{U}^*(\gamma,T_k) \hspace{-0.05cm} + \hspace{-0.35cm} \sum_{k=K+1}^{\infty} \hspace{-0.3cm} f(T_k;P_E)\Bar{U}_{w}(\gamma),  
\end{align*}
where $\Bar{U}_{w}(\gamma):=[\gamma H(X)+(1-\gamma)\mathbb{E}[X]]$ represents the worst case privacy-cost trade-off, in which $H(X)$ and $\mathbb{E}[X]$ are the entropy and expected value of the demand, respectively. Hence, we choose the minimum $K$ value that satisfies $\sum^{\infty}_{k=K+1}f(T_{k};P_E)\Bar{U}_w(\gamma)= (1-P_E)^{T_K}\Bar{U}_w(\gamma)<\epsilon$. We can find a finite $T_K$ satisfying this inequality for any $\epsilon > 0$.

\vspace{-0.4cm}
\section{A simple binary example}

\vspace{-0.35cm}
We consider a simple scenario with $(\mathcal{X}$,$\mathcal{Y})$=$\{0,1\}$, $\mathcal{E}$=$\{0,2\}$ and $\mathcal{B}$=$\{0,1,2\}$. We emphasize that obtaining numerical results for larger alphabets is challenging as the belief grows with the state space, and so does the computational complexity, also due to the quantization of the belief. For simplicity, demand and energy generation processes are assumed to be i.i.d. with Bernoulli $P_X$=$0.5$ and $P_E \in [0,1]$, respectively. Extensions to Markovian $E_t$ process is straightforward for TP and BCP; however, the MDP formulation requires including $E_t$ in the state, and updating the belief accordingly. We consider a privacy-cost trade-off weight of $\gamma$=$0.5$.

\begin{figure}[h]
\centering
\includegraphics[width=7.5cm,height=5.2cm]{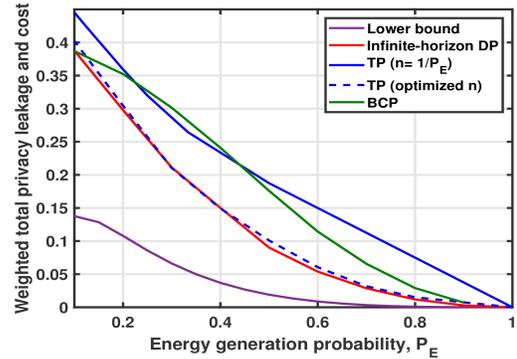}
\vspace{-0.4cm}
\caption{Privacy-cost trade-off of the lower bound, TP, BCP and infinite-horizon MDP w.r.t. $P_E$ for $\gamma$=$0.5$ and $P_X$=$0.5$.} 
\label{fig:result2}
\end{figure}

The weighted total privacy leakage and energy cost for TP, BCP and infinite-horizon MDP are depicted in Fig. \ref{fig:result2}, together with the lower bound. The average weighted cost decreases with $P_E$, since the demand can be mostly supplied by the RES, decreasing both the cost and leakage. The lower bound is obtained from (\ref{eqn:LB}) evaluated over a sufficiently long $T$. While the lower bound is not tight in general, it also shows us the value of predicting the energy generation instances for optimizing the privacy and cost. Two plots of TP are obtained corresponding to different horizons. For the first TP plot, the finite-horizon is set to be $n$=$\frac{1}{P_E}$. Since TP leads to full information leakage when energy arrives later than the set horizon, this approach has a higher privacy-cost trade-off compared to the infinite-horizon DP solution of the original problem. For the second TP plot, for each $P_E$ value, the best horizon value is selected by searching over the set $n=[1:15]$. We observed that, the optimal fixed horizon is typically longer than $\frac{1}{P_E}$, which reduces the probability of full leakage. Interestingly, the performance of TP with optimized yet fixed horizon follows that of the infinite-horizon MDP solution very closely. We remark here that the curve obtained for the infinite-horizon MDP solution is an approximation as well, due to the quantization of the belief. Finally, we observe that the performance of the BCP scheme can outperform that of fixed horizon TP policy for high $P_E$ values.

\vspace{-0.36cm}

\vspace{-0.8cm}
\section{CONCLUSIONS}

\vspace{-0.4cm}
We have studied the privacy-cost trade-off in a SM system equipped with both RB and RES. Motivated by the episodic nature of the problem,  we proposed a low-complexity TP policy to solve this infinite-horizon problem by solving simplified finite-horizon problems with only RB and no RES. We also proposed the BCP policy, whose actions depend only on the demand. We numerically showed for a binary example that the fixed-horizon policy that ignores the RES process can achieve a near-optimal performance. As a future work, we will try to quantify/bound the gap between the two policies.

\vfill\pagebreak


\bibliographystyle{IEEEbib}
\bibliography{Reference}

\end{document}